\def\cA{{\cal A}}
\def\co{{\cal O}}

\def\plushc{\ + h.c.}

\newcommand{\I}{{\rm i}}

\newcommand{\re}{{\rm Re}}
\newcommand{\im}{{\rm Im}}

\newcommand{\diag}{{\rm diag}}
\newcommand{\cL}{{\cal L}}
\newcommand{\cD}{{\cal D}}
\newcommand{\cF}{{\cal F}}

\newcommand{\nn}{\nonumber}
\newcommand{\al}[1]{\begin{eqnarray}#1\end{eqnarray}}
\newcommand{\eq}[1]{\begin{equation}#1\end{equation}}

\def\be{\begin{equation}}
\def\ee{\end{equation}}
\def\bea{\begin{eqnarray}}
\def\eea{\end{eqnarray}}

\documentstyle[epsf, 12pt]{article}
\begin{document}

\thispagestyle{empty}
\begin{flushright}
JHU-TIPAC-99011\\
hep-th/9912080\\
\end{flushright}

\bigskip\bigskip\begin{center} {\bf
\LARGE{\bf Partial Supersymmetry
Breaking from Five Dimensions} }
\end{center} \vskip 1.0truecm

\centerline{\bf Richard Altendorfer}

\vskip5mm
\centerline{\it Department of Physics and Astronomy}
\centerline{\it The Johns Hopkins University}
\centerline{\it 3400 North Charles Street}
\centerline{\it Baltimore, MD 21218, U.S.A.}
\vskip5mm

\bigskip \nopagebreak \begin{abstract}
\noindent Theories of partial supersymmetry breaking $N=2 \rightarrow N=1$
in four dimensions are derived by coupling the $N=2$  massless gravitino multiplet
to $N=2$ supergravity in five dimensions and performing a generalized
dimensional reduction on $S^1/Z_2$ with the Scherk-Schwarz mechanism.
These theories agree with results that were previously derived from four dimensions.
\end{abstract}

\newpage\setcounter{page}1

\section{Introduction}

Most supersymmetry phenomenology is based on the minimal
supersymmetric standard model, which is assumed to be
the low energy limit of a more fundamental theory like string theory or M-theory.
Those fundamental theories are formulated in higher space-time dimensions
and possess an extended number of supersymmetries which have to be
spontaneously broken to arrive at a phenomenologically viable $N=1$
model in four dimensions. Hence dimensional reduction and  partial supersymmetry
breaking are both present in this scenario.
A technical tool that satisfies both features simultaneously
is provided by the Scherk-Schwarz mechanism \cite{ss}
(also called generalized dimensional reduction), where nontrivial 
boundary conditions for the fields along the compactified dimensions
can lead to spontaneously broken gauge and space-time symmetries. 
In this paper the simplest setup to investigate partial supersymmetry
breaking by generalized dimensional reduction is used, namely
the partial breaking of $N=2\rightarrow N=1$ in four dimensions
coming from a theory in five dimensions. The focus is on
the superHiggs effect, where Goldstone fermions from the broken supersymmetries
become the longitudinal components of massive gravitinos. 

The physics that underlies this superHiggs effect for the spontaneous
breaking of  $N=2$ supersymmetry
to $N=1$ was investigated in
a flat Minkowski background \cite{np}
as well as in an anti-de Sitter  (AdS) background \cite{ab}, where
theories were constructed that describe partial supersymmetry breaking
in a model-independent approach with a minimal field content
motivated by the superHiggs effect. 
In this paper these theories of partial $N=2 \rightarrow N=1$
supersymmetry breaking in four dimensions will be reproduced by 
compactifying the massless $N=2$ $D=5$ massless gravitino multiplet\footnote{
Here, $N=2$ refers to the lowest supersymmetry in five dimensions, corresponding
to two symplectic Majorana spinors, which are equivalent to one Dirac spinor.}
on the orbifold $S^1/Z_2$ and using the Scherk-Schwarz mechanism \cite{ss} to introduce
a symmetry-breaking mass parameter.\footnote{In Ref. \cite{alledged} such a combination of 
orbifold projection and Scherk-Schwarz compactification was introduced to derive string
theories with spontaneously broken supersymmetry.}
In order to exhibit the superHiggs effect,
this procedure must be carried out in a supergravity background.

This derivation serves several purposes:
The starting point of the investigation is the massless 
$N=2$ $D=5$ gravitino multiplet 
Noether-coupled to $N=2$ $D=5$ supergravity which leads to 
pure $N=4$ $D=5$ supergravity (without matter). This is a rather simple
theory that allows for easy extensions such as embedding into a higher $N$
theory or introducing matter couplings.
Second, the theories obtained from five dimensions are already
out of unitary gauge, so that
the symmetry breaking mass parameter $m$ can be set 
to zero with impunity. In particular, no singular St\"uckelberg
redefinitions (containing ${1/m}$-terms as in Ref. \cite{np}) are required. 

What other motivation apart from simplicity is there 
to look at a theory in five dimensions?
It seems natural to derive
a massive 4-dimensional theory from a massless theory
in five dimensions, because the degrees of freedom of
a massless $D=5$ vector field (3) and the degrees of
freedom of a massless symplectic $D=5$ Majorana spinor
(4) match those of their massive counterparts in $D=4$.
This transition can be easily implemented by a 
Scherk-Schwarz compactification on $S^1$, where it can be seen
that the fifth components of a symplectic  gravitino $\Psi^i_{\alpha M}$
and a complex vector $A_M$, $\Psi^i_{\alpha \dot 4}$ and  $A_{\dot 4}$, provide
the longitudinal components of the massive 4-dimensional
fields $\Psi^i_{\alpha m}$ and  $A_m$.\footnote{In this paper,
$i,j$... denote symplectic indices, $M,N$... are 5-dimensional world indices 
and $m,n$... are 4-dimensional world indices. A dotted numerical index stands for a
world index, to be distinguished from undotted Lorentz indices.}

A natural candidate for partial supersymmetry breaking in four 
dimensions effected by Scherk-Schwarz compactification on $S^1$ would be
pure $N=2$ $D=5$ supergravity \cite{crem, chams}, in particular since the degrees
of freedom exactly match those of one of the minimal cases
(with two vectors) considered in Ref. \cite{np}. Unfortunately,
Scherk-Schwarz compactification applied to this theory 
 breaks both supersymmetries. In fact,
any Scherk-Schwarz compactification of $N$-extended supergravity on $S^1$ breaks 
an even number of supersymmetries, since the Scherk-Schwarz
generator must be a generator of $USp(N)$ \cite{d5n8}. This can be circumvented
by projecting out half of the states by compactifying on
the orbifold $S^1/Z_2$. In this paper, the length of the interval
$S^1/Z_2$ is assumed to be of the order of the Planck length; hence
only the massless Kaluza-Klein modes (possibly lifted by
the Scherk-Schwarz mass parameter) are retained.

The outline of the paper is as follows: In the second section
the massless $N=2$ $D=5$ gravitino multiplet will be
compactified on $S^1/Z_2$, yielding a massive $N=1$ gravitino 
multiplet in four dimensions together with a spontaneously
broken fermionic gauge symmetry. In the third section, the
massless $N=2$ $D=5$ gravitino multiplet will be Noether-coupled
to $N=2$ $D=5$ supergravity thus yielding $N=4$ supergravity before the dimensional reduction
is carried out. The corresponding 4-dimensional theory 
is shown to coincide with theories exhibiting partial supersymmetry
breaking which were derived from four
dimensions \cite{zin, zinovev, fgp-local}.\footnote{The geometric construction
of \cite{fgp-local} could be generalized to accommodate an arbitrary 
$N=2$ matter and gauge content \cite{pesando}
and to give masses to light mirror fermions \cite{mirror}. String models exhibiting the
mass spectrum of partially broken supersymmetry were constructed in Refs. \cite{toni}.}

\section{Generalized compactification of the massless $N=2$ $D=5$ gravitino multiplet \label{section2}} 

The massless $N=2$ $D=5$ gravitino multiplet contains two symplectic Majorana gravitinos
$\Psi_{iM}$, two symplectic Majorana spinors $\Lambda_i$ and four real vector fields. 
Its Lagrangian and supersymmetry transformations can be conveniently
obtained by truncation of the $N=4$ $D=5$ supergravity constructed in Ref. \cite{d5n4}. The conventions
for the 5-dimensional Dirac algebra and the symplectic geometry are presented in the Appendix.

The five abelian vector fields $A^{ij}_M$ $i,j \in \{ 1,...,4 \}$ in the 5-dimensional representation of $USp(4)$ 
which are part of the spectrum of $N=4$ $D=5$ supergravity satisfy
the reality condition \cite{crem}

$$
A^{ij}_M = (A_{Mij})^* \ ,
$$ 
and are parametrized by
\eq{
A_{Mij}= \pmatrix{0 & A_M & \bar B_M & \bar C_M  \cr
                  -A_M & 0 & -C_M & B_M \cr
                  -\bar B_M & C_M & 0 & -A_M \cr
                  -\bar C_M & -B_M & A_M & 0 \cr } \ , \label{fieldstr}
}
where $A_M$ is real and $B_M$ and $C_M$ are complex.

Truncation of the $N=4$ $D=5$ supergravity to the massless $N=2$ gravitino multiplet
requires $A_M=0$. The remaining two complex vectors can be written as

$$
A_{Mia} =  \pmatrix{-\bar B_M & C_M \cr -\bar C_M & -B_M \cr} \qquad a,i \in \{ 1,2 \}\ ,
$$
which corresponds to a decomposition $USp(4) \rightarrow USp(2) \otimes USp(2) \cong SU(2) \otimes SU(2) $. 
The field $A_M$ now carries two different symplectic indices $i$ and $a$, indicating the two independent 
$SU(2)$'s that it is charged under.
Here, 
$$
\Omega_{ij} = \Omega_{ab} = \pmatrix{ 0 & 1 \cr -1 & 0 \cr }\ .
$$
The Lagrangian
is given by
\eq{
\cL = {\I\over 2}\bar\Psi_{iM} \Gamma^{MNO}\partial_N \Psi^i_O -{\I\over 2}\bar\Lambda_i \Gamma^M\partial_M \Lambda^i 
    + {1\over 8} {{F_{MN}}^i}_a {{{F}^{MNa}}_i}\ , \label{notdual}
} 
and the supersymmetry transformations are
\al{
\delta_\Xi \Psi_M^i &=& {1\over 6}{{F_{NO}}^i}_a({\Gamma_M}^{NO} + 2 \delta_M^{[ N}\Gamma^{O ]})\Xi^a \nn \\ 
\delta_\Xi {A_{Ma}}^i &=&  -{2\I\over \sqrt{3}}\bar\Xi_a\Gamma_M \Lambda^{i} - {2\I}\bar\Xi_a\Psi^{i}_M \nn\\
\delta_\Xi \Lambda^i &=& -{1\over 2\sqrt{3}}{{F_{MN}}^i}_a\Gamma^{MN}\Xi^a \ ,
}
where ${{F_{MN}}^i}_a = \partial_{[M}{{A_{N]}}^i}_a$.
In order to implement the Scherk-Schwarz mechanism on the orbifold $S^1/Z_2$, the $Z_2$ 
transformation and the generator $T$ of a global symmetry of the 5-dimensional theory used
for the Scherk-Schwarz mechanism must satisfy \cite{pomarol}
\eq{
Z_2 e^{\I m T x^4} = e^{\I m T (-x^4)} Z_2 \ \Leftrightarrow\ \{ Z_2, T \} = 0 \ . \label{z2}
}
In addition, the parity operation must be part of the discrete symmetries of the 5-dimensional theory.
The effect of the parity operator $Z_2 = \pm\pmatrix{1 & 0 \cr 0 & -1 \cr}\otimes \I\Gamma^4$
on the fields is listed in Tables 1 and 2. 
With the choice $T = \sigma^2 \in su(2)$ and the transformation
$$\Phi^i \rightarrow {(e^{\I m \sigma^2 x^4})^{i}}_j \Phi^j$$
for a generic field $\Phi^i$ having a symplectic $i$ (not $a$) index,
the generalized dimensional reduction of the massless gravitino multiplet is completely specified.
\begin{table}[t]
\centerline{
\begin{tabular}{|c|c|c|c|c|c|} \cline{1-6}
  & $ \lambda^i $ & $\psi^i_{m}$ & $\psi^i_{4}$ & $\xi^a$ & $i/a$\\ \hline
{Parity} & 1 & -1 & 1 & 1 & 1 \\ \cline{2-6}
  & -1 & 1 & -1 & -1 & 2\\ \hline
\end{tabular}
}
\caption{Fermionic parity assignment of $D=5$ $N=2$ gravitino multiplet in terms of $D=4$ Weyl spinors.}
\end{table}

\begin{table}[t]
\centerline{
\begin{tabular}{|l|c|c|c|c|} \cline{1-5}
 & $B_m$ & $B_4$ & $C_m$ & $C_4$\\ \hline
Parity & -1 & 1 & 1 & -1 \\ \hline
\end{tabular}
}
\caption{Bosonic parity assignment of $D=5$ $N=2$ gravitino multiplet in terms of $D=4$ fields.}
\end{table}
The limit $l \rightarrow 0$ of the length $l$ of the interval $S^1/Z_2$ is
implicitly understood, so that only the lowest even Fourier-modes of the fields
which depend only on the first four space-time co-ordinates $x^0, ..., x^3$ (zero-modes)
need to be retained.

To simplify the identification of the physical fields in the 4-dimensional theory,
the following redefinitions are necessary to diagonalize and normalize 
the fermionic kinetic terms:
\eq{
\Psi^i_{m} \rightarrow  \Psi^i_{m} + {1\over 2}\Gamma_m\Gamma^4\Psi^i_{4} \ , \label{redef}
}
and $\nu^i = \sqrt{3 / 2}\psi^i_{4}$. Note that in (\ref{redef}) no terms $\sim 1/m \partial_m \Psi^i_{4}$
as in Ref. \cite{np} occur; in a gravitational background, the covariantized form of this term would not
leave the gravitino kinetic term invariant and would therefore induce terms $\sim 1/m$ in the Lagrangian.

With additional ``chiral'' redefinitions $\Lambda^i \rightarrow -\Gamma^4\Lambda^i$ and $\Xi^a \rightarrow \Gamma^4\Xi^a$
the Lagrangian becomes
\al{
\cL_4 & = &  \epsilon^{mnpq} \bar \psi_{2m}
  \bar \sigma_n \partial_p \psi^2_q
 - \I \bar \lambda_1 \bar \sigma^m\partial_m \lambda^1 
 - \I \bar \nu_1 \bar \sigma^m\partial_m \nu^1  \nonumber \\
& & - {1 \over 4} C_{mn} \bar C^{mn} 
- {1\over 2}\cD_m B_4 \cD^m \bar B_4
    \nonumber \\
& & - \sqrt{{3 \over 2}} m  ( \I\psi^2_m  \sigma^m
   \bar \nu_1 \plushc) - m (\nu^1 \nu^1 \plushc) \nonumber \\ 
& & +{1\over 2}m\lambda^1\lambda^1  - m\psi^2_m \sigma^{mn} \psi^2_n \plushc 
}
with supersymmetry transformations
\al{
\delta_\xi C_m &=& \I{2\over\sqrt{3}}\xi^1\sigma_m\bar\lambda_1 + 2\xi^1\psi^2_m -\I
\sqrt{2\over 3}\xi^1\sigma_m\bar\nu_1 \nn \\
\delta_\xi B_4 &=& {2\over\sqrt{3}}\xi^1\lambda^1 +  {2\sqrt{2}\over\sqrt{3}}\xi^1\nu^1 \nn \\
\delta_\xi \lambda^1 &=& {1\over\sqrt{3}}\bar C_{mn}\sigma^{mn}\xi^1 + {\I\over\sqrt{3}}\cD_mB_4\sigma^m\bar\xi_1 \nn\\
\delta_\xi \psi^2_m &=& - {\I\over 2}C_{+mn}\sigma^n\bar\xi_1 + \cD_m\bar B_4\xi^1 \nn \\
\delta_\xi \nu^1 &=& - {1\over\sqrt{6}}\bar C_{mn}\sigma^{mn}\xi^1 + \I\sqrt{2\over 3}\cD_m B_4 \sigma^m\bar\xi_1 \ ,\nn
}
where $\cD_m B_4 = \partial_m B_4 - m C_m$ and $C_{+mn}=\partial_{[m}C_{n]}+{\I\over 2}\epsilon_{mnrs}\partial^{[r}C^{s]}$.
The expression for the gauge invariant derivative $\cD_m$ shows that the complex scalar $B_4$ is the Goldstone boson
of a spontaneously broken abelian gauge symmetry mediated by the vector $C_m$. The spinor $\nu^1$ 
is the longitudinal component of
the massive gravitino $\psi^2_m$. The massless five-dimensional theory also contains a fermionic gauge symmetry:
$$
\delta_\Theta \Psi^i_M = {2\over \kappa}\partial_M \Theta^i \ .
$$
A generalized Scherk-Schwarz compactification leads to a spontaneously broken fermionic symmetry:
\al{
\delta_\theta \psi^2_m &=& \I {m\over \kappa} \sigma_m \bar\theta_2     \nn\\
\delta_\theta \nu^1    &=& \sqrt{6} {m\over \kappa} \theta^2 \ . \label{fermsym}
}
Therefore, $\nu^1$ can also be interpreted as the Goldstino for this broken symmetry.

The Lagrangian for the massless gravitino multiplet (\ref{notdual}) is not unique.
In flat 5-dimensional space-time a massless vector $B_M$ is dual\footnote{This is not true for a 5-dimensional
AdS background, where antisymmetric tensors satisfy additional self-duality conditions \cite{guma}.} 
to an antisymmetric tensor $G_{MN}$ 
\eq{
\partial_{[M}B_{N]}= {1\over 2}\epsilon_{MNOPQ} \partial^OG^{PQ} + \cdots , \label{dual}
}
where the dots stand for contributions from interaction terms. 
So all four real vectors of the massless $N=2$ gravitino multiplet
can be dualized and then dimensionally reduced to four dimensions. 
The implementation of the Scherk-Schwarz mechanism on $S^1/Z_2$ requires that fields of the same index structure
come in pairs, so that one of them can be projected out by the $Z_2$ reflection whereas the other becomes a massive
field in the 4-dimensional theory. 
If both $B_M$ and
$C_M$ are dualized, the field strength ${{F_{MN}}^i}_a$ is simply replaced by its dual in the Lagrangian
and the supersymmetry transformations and the automorphism group
decomposes as before as $USp(4) \rightarrow SU(2) \otimes SU(2)$.

The other possibility is the dualization of only two vectors, which must be the real or imaginary parts
of the vectors $B_M=B_M^R+\I B^I_M$ and $C_M=C_M^R+\I C^I_M$. 
Otherwise, either the complex vector or the complex antisymmetric tensor
are completely projected out. Here, the imaginary parts are chosen to be dualized: $B^I_M\rightarrow 
B^I_{MN}$ and $C^I_M\rightarrow  C^I_{MN}$.
This corresponds to the decomposition 
\al{
{{F_{MN}}^i}_a &=& \re({{F_{MN}}^i}_a) + \I\im({{F_{MN}}^i}_a) \nn\\
          &\rightarrow&  {{F^R_{MN}}^i}_a + \I {{v_{MN}}^i}_a  \ , \nn
}
where ${{v_{MN}}^i}_a =1/2\epsilon_{MNOPQ}
\partial^O{{G^{PQ}}^i}_a$ and ${{G_{MN}}^i}_a = \pmatrix{- C^I_{MN} & B^I_{MN} \cr B^I_{MN} & C^I_{MN} \cr}$. To conserve this structure, the $SU(2) \otimes SU(2)$ transformations
must be restricted to those which do not mix real and imaginary parts. Therefore only the generator
$\sigma^2$ is allowed, which corresponds to a decomposition $SU(2) \otimes SU(2) \rightarrow 
U(1)\otimes U(1)$.

Dualization of the imaginary parts of the complex vectors $B_M$ and $C_M$ yields the Lagrangian
\al{
\cL &=&{\I\over 2}\bar\Psi_{iM} \Gamma^{MNO}\partial_N \Psi^i_O -{\I\over 2}\bar\chi_i \Gamma^M\partial_M \chi^i\nn \\
    & & +{1\over 8} {{F^R_{MN}}^i}_a {{{F}^{RMNa}}_i} + {1\over 8} {{v_{MN}}^i}_a {{{v}^{MNa}}_i} \ ,
} 
and the supersymmetry transformations are
\al{
\delta_\Xi \Psi_M^i &=& {1\over 6}{{F^R_{NO}}^i}_a({\Gamma_M}^{NO} + 2 \delta_M^{[ N}\Gamma^{O ]})\Xi^a +
                        {\I\over 6}{{v_{NO}}^i}_a({\Gamma_M}^{NO} + 2 \delta_M^{[ N}\Gamma^{O ]})\Xi^a \nn \\ 
\delta_\Xi {A^R_{Ma}}^i &=&  -{\I\over \sqrt{3}}\bar\Xi_a\Gamma_M \Lambda^{i} - {\I}\bar\Xi_a\Psi^{i}_M \plushc\nn\\
\delta_\Xi {{G_{MN}}_a}^i &=&  {1\over \sqrt{3}}\bar\Xi_a\Gamma_{MN} \chi^{i} - \bar\Xi_a\Gamma_{[M }\Psi^{i}_{ N]}\plushc\nn \\
\delta_\Xi \Lambda^i &=& -{1\over 2\sqrt{3}}{{F^R_{MN}}^i}_a\Gamma^{MN}\Xi^a -{\I\over 2\sqrt{3}}{{v_{MN}}^i}_a\Gamma^{MN}\Xi^a \ .
}
The parities of the antisymmetric tensors in terms  of 4-dimensional fields are determined by the 
parities of the dual vectors and the dualization relation (\ref{dual}) and are listed in Table \ref{table3}. 
\begin{table}[t]
\centerline{
\begin{tabular}{|l|c|c|c|c|c|c|c|c|} \cline{1-9}
 & $B^R_m$ & $B^R_4$ & $C^R_m$ & $C^R_4$ & $B^I_{mn}$ & $B^I_{m4}$ & $C^I_{mn}$ & $C^I_{m4}$ \\ \hline
Parity & -1 & 1 & 1 & -1 & 1 & -1 & -1 & 1 \\ \hline
\end{tabular}
}
\caption{Bosonic parity assignment of the dualized $D=5$ $N=2$ gravitino multiplet in terms of $D=4$ fields.}\label{table3}
\end{table}
Performing the same procedure and redefinitions (\ref{redef}) as before, the following Lagrangian is 
obtained:
\al{
\cL_4 & = &  \epsilon^{mnpq} \bar \psi_{2m}
  \bar \sigma_n \partial_p \psi^2_q
 - \I \bar \lambda_1 \bar \sigma^m\partial_m \lambda^1 
 - \I \bar \nu_1 \bar \sigma^m\partial_m \nu^1  \nonumber \\
& & - {1 \over 4} C^R_{mn} C^{Rmn} 
- {1 \over 4}\cF_{mn}^{C^I_4}\cF^{C^I_4mn} + {1 \over 4}v^{B^I}_m v^{B^Im} - {1\over 2}\cD_m B^R_4 \cD^m  B^R_4
    \nonumber \\
& & - \sqrt{{3 \over 2}} m  ( \I\psi^2_m  \sigma^m
   \bar \nu_1 \plushc) - m (\nu^1 \nu^1 \plushc) \nonumber \\ 
& & +{1\over 2}m\lambda^1\lambda^1 - m\psi^2_m \sigma^{mn} \psi^2_n \plushc 
}
with supersymmetry transformations
\al{
\delta_\xi C^R_m &=& \I{1\over\sqrt{3}}\xi^1\sigma_m\bar\lambda_1 + \xi^1\psi^2_m -\I
{1\over \sqrt{6}}\xi^1\sigma_m\bar\nu_1 \plushc \nn \\
\delta_\xi B^R_4 &=& {1\over\sqrt{3}}\xi^1\lambda^1 + {\sqrt{2\over 3}}\xi^1\nu^1 \plushc\nn \\
\delta_\xi B^I_{mn} &=& -{2\over\sqrt{3}}\xi^1\sigma_{mn}\lambda^1 -2\sqrt{2\over 3}\xi^1\sigma_{mn}\nu^1 
 -\I\xi^1\sigma_{[m}\bar\psi_{2n]} \plushc \nn \\
\delta_\xi C^I_{m4} &=& -{\I\over\sqrt{3}}\xi^1\sigma_m\bar\lambda_1 + {\I\over\sqrt{6}}\xi^1\sigma_m\bar\nu_1 
+ \xi^1\psi^2_m \plushc \nn \\
\delta_\xi \lambda^1 &=& {1\over\sqrt{3}} C^R_{mn}\sigma^{mn}\xi^1 + {\I\over\sqrt{3}}\cD_mB^R_4\sigma^m\bar\xi_1 
       - {1\over\sqrt{3}}\cF_{mn}^{C^I_4} \sigma^{mn}\xi^1 + {1\over\sqrt{3}}v^{B^I}_m\sigma^m\bar\xi_1    \nn\\
\delta_\xi \nu^1 &=& - {1\over\sqrt{6}} C^R_{mn}\sigma^{mn}\xi^1 + \I\sqrt{2\over 3}\cD_m B^R_4 \sigma^m\bar\xi_1 
   + {1\over\sqrt{6}}\cF_{mn}^{C^I_4} \sigma^{mn}\xi^1 + {\sqrt{2\over 3}}v^{B^I}_m\sigma^m\bar\xi_1  \nn \\
\delta_\xi \psi^2_m &=& - {\I\over 2}C^R_{+mn}\sigma^n\bar\xi_1 + \cD_m B^R_4\xi^1  
  -{\I\over 2}\cF_{+mn}^{C^I_4}\sigma^n\bar\xi_1 +\I v^{B^I}_m \xi^1 \ , \nn 
}
where $\cD_m B^R_4 = \partial_m B^R_4 - m C^R_m$, $\cF_{mn}^{C^I_4} = \partial_{[m}C^I_{n]4} - m B^I_{mn}$ and $v^{B^I}_m = 1/2\epsilon_{mnop}\partial^nB^{Iop}$. In the dual case, 
the vector $C^I_{m4}$ becomes the Goldstone boson of the spontaneously
broken gauge symmetry mediated by the antisymmetric tensor $B^I_{mn}$. This is the 
dual Higgs mechanism investigated in Ref. \cite{quevtrug}.
As before, the 5-dimensional theory has a fermionic gauge symmetry which is spontaneously broken
upon generalized dimensional reduction (\ref{fermsym}).
The generalized dimensional reduction of the massless gravitino multiplet with four antisymmetric tensors is 
analogous to the case illustrated above and will not be presented here.

These 4-dimensional theories correspond exactly, up to field relabelings,
to the dual versions of the massive $N=1$ gravitino multiplet (out of unitary gauge)
before being coupled to gravity \cite{np}. In order to exhibit the superHiggs 
effect for partial supersymmetry breaking, the above multiplets must be coupled
to gravity and the fermionic gauge symmetry must be promoted to a local supersymmetry. 
This will be addressed in the next section. 

\section{Generalized dimensional reduction of pure $N=4$ $D=5$ supergravity}

The complete theory of partially broken supergravity in four dimensions to all orders in the fields 
should now be obtainable by Noether-coupling the massless gravitino multiplet 
to pure $N=2$ $D=5$ supergravity \cite{crem, chams, d5n2} and performing a generalized dimensional
reduction. 
Since consistency requires that the fermionic gauge symmetries
of the massless gravitinos become local supersymmetries, the resulting theory must be $N=4$ supersymmetric.
Therefore, the Scherk-Schwarz mechanism should be applied to $N=4$ $D=5$ supergravity.
The pure five-dimensional $N=4$ supergravity was shown to be derivable 
as a consistent truncation of the maximal $N=8$ supergravity \cite{crem}. The explicit form of 
the $N=4$ supergravity Lagrangian and its supersymmetry transformations was given
in Ref. \cite{d5n4}.

For the sake of clarity representative parts of the result of \cite{d5n4} are
reproduced\footnote{Here, the vector in the singlet representation of $USp(4)$, $B_\mu$,  from Ref. \cite{d5n4} 
has been renamed to $G_M$ and the vector fields in the 5-dimensional representation of $USp(4)$
are parametrized as in (\ref{fieldstr}). Also, the 5-dimensional spinors $\chi^i$ have been renamed to $\Lambda^i$. 
The following redefinitions were performed: $\Gamma^A\rightarrow -\I\Gamma^A$, 
$A_{Mij}\rightarrow 1/2 A_{Mij}$, and $\Xi^i\rightarrow 2\Xi^i$.}. 
$N=4$ supergravity has a $USp(4)$ symmetry inherited from the automorphism group of
the supersymmetry algebra
and its field content together with a consistent parity assignment for inversion 
of the fifth co-ordinate is given in Tables 4 and 5. 
The Lagrangian is given up to terms of order $\co(\kappa)\co({\rm fermions})$\footnote{The Lagrangian derived in  Ref. \cite{d5n4}
includes such terms up to four-fermion terms, but they are not illustrative in the context of partial supersymmetry breaking.} 
and the 
supersymmetry transformations are given up to three-fermion terms. 
The Lagrangian reads
\begin{table}[t]
\centerline{
\begin{tabular}{|c|c|c|c|c|c|} \cline{1-6}
  & $ \lambda^i $ & $\psi^i_{m}$ & $\psi^i_{\dot 4}$ & $\xi^i$ & $i$ \\ \hline
Parity &  -1 & 1 & -1 & \multicolumn{1}{||c||}{1} & 1 \\ \cline{2-6}
 & 1 & -1 & 1 & \multicolumn{1}{||c||}{ -1} & 2 \\ \cline{2-6}
 & \multicolumn{1}{||c|}{ 1} & -1 & \multicolumn{1}{c||}{ 1} & -1 & 3 \\ \cline{2-6}
 & \multicolumn{1}{||c|}{ -1} & 1 & \multicolumn{1}{c||}{-1} & 1 & 4 \\ \hline
\end{tabular}
}
\caption{Fermionic fields and parities of $D=5$ $N=4$ supergravity in terms of $D=4$ Weyl spinors.}
\end{table}
\begin{table}[t]
\centerline{
\begin{tabular}{|l|c|c|c|c|c|c|c|c||c|c|c|c||} \cline{1-13}
 & ${e^a}_{m}$ & ${e^4}_{m}$ & ${e^4}_{\dot 4}$ & $G_m$ & $G_{\dot 4}$ & $\sigma$ & $A_m$ & $A_{\dot 4}$ & $B_m$ & $B_{\dot 4}$ & $C_m$ & $C_{\dot 4}$\\ \hline
Parity & 1 & -1 & 1 & -1 & 1 & 1 & -1 & 1 & -1 & 1 & 1 & -1 \\ \cline{2-13}
       & \multicolumn{5}{|c|}{$N=2$ supergravity} & \multicolumn{3}{|c||}{vector mult.} & 
\multicolumn{4}{|c||}{gravitino mult.} \\ \hline
\end{tabular}
}
\caption{Bosonic fields and parities of $D=5$ $N=4$ supergravity in terms of $D=4$ fields.}
\end{table}
\al{
\kappa e^{-1}\cL &=& -\ {1 \over 2
\kappa^2} {\cal R} + {\I\over 2}\bar\Psi_{iM} \Gamma^{MNO}D_N \Psi^i_O 
- {\I\over 2}\bar\Lambda_i \Gamma^M D_M \Lambda^i \nn \\
    & & -{1\over 16} e^{{2\over\sqrt{3}}\kappa\sigma}F_{MN}^{ij} {F}^{MN}_{ij} 
-{1\over 4} e^{-{4\over\sqrt{3}}\kappa\sigma}G_{MN} {G}^{MN} -{1\over 2}\partial_M\sigma\partial^M \sigma \nn \\
& & + {\kappa\over 16\sqrt{2}}e^{-1}\epsilon^{MNOPQ}F_{MN}^{ij} {F}_{OPij} G_Q + \co(\kappa)\co({\rm fermions})\label{five} \ ,
}
and the supersymmetry transformations are
\al{
\delta_\Xi \Psi_{Mi} &=& {2\over\kappa}D_M \Xi_i-{1\over 6}e^{{1\over\sqrt{3}}\kappa\sigma}F_{NOij}({\Gamma_M}^{NO} + 2 \delta_M^{[ N}\Gamma^{O ]})\Xi^j \nn \\
                    & & -{1\over 6\sqrt{2}}e^{-{2\over\sqrt{3}}\kappa\sigma}G_{NO}({\Gamma_M}^{NO} + 2 \delta_M^{[ N}\Gamma^{O ]})\Xi_i 
                        + (\hbox{3-fermion terms}) \nn \\
\delta_\Xi {A_{M}}^{ij} &=&  {\I\over \sqrt{3}}e^{-{1\over\sqrt{3}}\kappa\sigma}(2\bar\Xi^{[i}\Gamma_M \Lambda^{j]} 
- \Omega^{ij}\bar\Xi_k\Gamma_M \Lambda^k) \nn \\
& & -\I e^{-{1\over\sqrt{3}}\kappa\sigma}(2\bar\Xi^{[i}\Psi^{j]}_M - \Omega^{ij}\bar\Xi_k\Psi^k_M) \nn\\
\delta_\Xi \Lambda_i &=& -\Gamma^M\Xi_i\partial_M\sigma - {1\over 2\sqrt{3}}e^{{1\over\sqrt{3}}\kappa\sigma}F_{MNij}\Gamma^{MN}\Xi^j \nn \\
& &+{1\over \sqrt{6}}e^{{2\over\sqrt{3}}\kappa\sigma}G_{MN}\Gamma^{MN}\Xi_i + (\hbox{3-fermion terms}) \nn \\
\delta_\Xi G_M &=&  \I\sqrt{2\over 3}e^{{2\over\sqrt{3}}\kappa\sigma}\bar\Xi_i\Gamma_M \Lambda^i 
+ {\I\over\sqrt{2}}e^{{2\over\sqrt{3}}\kappa\sigma}\bar\Xi_i\Psi^i_M  \nn \\
\delta_\Xi \sigma &=& -\I \bar\Xi_i\Lambda^i \nn \\
\delta_\Xi {e^A}_M &=& \I \kappa\bar\Xi_i \Gamma^A \Psi^i_M \label{fivetrafo} \ .
}
The fields and supersymmetry parameters that are enclosed in double lines in Tables 4 and 5 are the ones coming from
the gravitino multiplet.  This field content shows that a consistent Noether-coupling of the gravitino
multiplet to $N=2$ supergravity necessitates the inclusion of a vector multiplet with fields 
($\lambda^1,\lambda^2, \sigma, A_m, A_{\dot 4}$).

From the parities in Tables 4 and 5 it is clear that the four dimensional theory contains in addition
to the fields of $N=2$ $D=4$ supergravity and the massive $N=1$ gravitino multiplet out of unitary gauge \cite{np} two
chiral multiplets ($\lambda^2, \psi^2_{\dot 4}, P, G_{\dot 4},\sigma,A_{\dot 4}$) where $e^{P}={e^4}_{\dot 4}$. 
This is the field content of one vector- and one hyper-multiplet coupled to $N=2$ supergravity.

The rest of this section is now devoted to showing that a generalized dimensional reduction of $N=4$ $D=5$ supergravity
on $S^1/Z_2$ with appropriately chosen $USp(4)$-generator for the Scherk-Schwarz mechanism
indeed leads to partially broken $N=2$ supergravity in four dimensions. First, those $USp(4)$-generators
which satisfy the condition (\ref{z2}) with $Z_2 = \pm \diag(1, -1, -1, 1)\otimes \I\Gamma^4$ must be singled out.
Out of $T_r$ $r \in \{1,...,10 \}$ (see Appendix),
the generators $T_s$ with $s \in \{ 1,2,4,5,8,10 \}$ satisfy (\ref{z2}). 

From the supersymmetry transformation of the gravitinos (\ref{fivetrafo})
\al{
\delta_\Xi \Psi_{iM} &=& {2\over\kappa}D_M\Xi_i + \cdots \nn\\
\rightarrow \qquad \delta_\Xi  {(e^{\I m T_s x^4})_{i}}^j  \Psi_{iM} &=& 
{2\over\kappa}D_M \left({(e^{\I m T_s x^4})_{i}}^j \Xi_j \right) + \cdots
}
it is obvious that the fifth component $\Psi_{i\dot 4}$ of the gravitinos can pick up
a constant shift $\sim m$. With the redefinition (\ref{redef}) the corresponding 
4-dimensional gravitino $\Psi_{im}$ will also pick up a shift, thus identifying the
broken supersymmetries. 

The choice $T_s$ with $s \in \{ 8,10 \}$
would break all supersymmetries. In order to match the Scherk-Schwarz mechanism in Sec. \ref{section2}, 
the generator
$T_5={\scriptscriptstyle\pmatrix{0 & 0 \cr 0 & \sigma^2 \cr}} $ will be used to implement the partial breaking of supersymmetry in the 4-dimensional theory. With $T_5$, both $\psi^3_{\dot 4}$ and $\psi^4_{\dot 4}$ acquire a shift under supersymmetry transformations
$$
\delta_\xi \pmatrix{\psi^3_{\dot 4} \cr \psi^4_{\dot 4} \cr} = 
{2\over\kappa}m\pmatrix{0 & 1 \cr -1 & 0 \cr}\pmatrix{\xi^3 \cr \xi^4} + \dots
$$
Since $\psi^4_{\dot 4}$ is projected out under the $Z_2$ reflection, only $\psi^3_{\dot 4}$ will be the Goldstino
of one spontaneously broken supersymmetry in the 4-dimensional theory. 

The dimensional reduction from $M_5\rightarrow M_4\times S^1/Z_2$ requires Weyl rescalings and field
redefinitions described in Refs. \cite{chams, weylre} in order to obtain canonically normalized diagonal kinetic terms.
They are simplified, however, because the vector ${e^4}_{m}$ from the f\"unfbein 
used for the Weyl rescaling is odd
under the $Z_2$ reflection. Explicitly the rescalings and redefinitions 
are: ${e^a}_{m} \rightarrow e^{-P/2}{e^a}_{m}$, 
$\xi^i \rightarrow e^{-P/4}\xi^i$, $\lambda^i \rightarrow -\I e^{P/4} \lambda^i$, 
$\psi^1_m \rightarrow e^{-P/4}\psi^1_m + \I/2\sigma_m\bar\psi_{2,\dot 4}e^{-3P/2}$, 
$\psi^4_m \rightarrow e^{-P/4}\psi^4_m - \I/2\sigma_m\bar\psi_{3,\dot 4}e^{-3P/2}$, 
$\psi^i_{\dot 4} \rightarrow e^{5P/4}\sqrt{2/3}\ \nu^i$.

The resulting 4-dimensional Lagrangian up to terms of order $\co(\kappa)\co({\rm fermions})$ 
in the conventions of \cite{wb} reads
\al{
&& e^{-1}\cL =\nn\\
&& -\ {1 \over 2 \kappa^2} {\cal R}
 + \epsilon^{pqrs} \bar \psi_{ip}
  \bar \sigma_q D_r \psi^i_{s} 
 - \I \bar \chi_i \bar \sigma^m D_m \chi^i 
 - \I \bar \Omega_i \bar \sigma^m D_m \Omega^i\nn\\
&& -\ {1 \over 4}e^{\sqrt{2}\kappa\varphi} \cA_{m n} \bar \cA^{m n}\! 
   -{1\over 2}\partial_m\varphi\partial^m\varphi
   -{1\over 2}e^{-2\sqrt{2}\kappa\varphi}\partial_m\pi\partial^m\pi
   -{1\over 2}\partial_m\tilde\varphi\partial^m\tilde\varphi
   -{1\over 2}e^{2\kappa\tilde\varphi}\cD_m\pi^a\cD^m\pi^a \nn \\
&& - m e^{\kappa(\tilde\varphi-{1\over\sqrt{2}}\varphi)}(-{\I\over \sqrt{2}} \psi^2_{m} \sigma^m
   \bar\Omega_1 
 + \I\psi^2_{m} \sigma^m
   \bar \chi_1 
- \sqrt{2} \Omega^1 \chi^1 
 +\ {1 \over 2} \chi^1 \chi^1 
+ \psi^2_{m} \sigma^{m n} \psi^2_{n} \plushc)\nn \\
&& +{\kappa\over 4\sqrt{2}}\pi\tilde\cA_{m n} \bar\cA^{m n} + \co(\kappa)\co({\rm fermions})\label{broke} \ ,
}
and the supersymmetry transformations up to three-fermion terms are
\al{
\delta_\xi {e^a}_m &=& -\I \kappa   \psi^i_{m}\sigma^a\bar\xi_i \plushc \nonumber \\
\delta_\xi \psi^i_{m} & = & {2 \over \kappa} D_m \xi^i
 - {1 \over 2}\epsilon^{ij}e^{\kappa{\varphi\over\sqrt{2}}} \cA_{+m n}\sigma^n \bar \xi_j
 + {\I \over \sqrt{2}}e^{-\sqrt{2}\kappa\varphi}\partial_m\pi\xi^i \nn \\
& & -\I e^{\kappa\tilde\varphi}\cD_m\pi^a{{\sigma^a}^i}_j\xi^j
 +\I {m\over \kappa}\delta^i_2 e^{\kappa(\tilde\varphi-{1\over\sqrt{2}}\varphi)}\sigma_m\bar\xi_2\nn \\
& & + (\hbox{3-fermion terms})\nn \\
\delta_\xi \cA_m &=& e^{-\kappa{\varphi\over\sqrt{2}}}(-2\I \epsilon_{i j} \psi^i_{m} \xi^j 
- \sqrt{2} \bar\Omega_i \bar \sigma_m \xi^i) \nonumber \\
\delta_\xi \Omega^i  &=&  -{\I \over \sqrt{2}}e^{\kappa{\varphi\over\sqrt{2}}} \bar\cA_{mn} 
\sigma^{mn} \xi^i -\I\epsilon^{i j}\sigma^m\bar\xi_j(\partial_m\varphi -\I e^{-\sqrt{2}\kappa\varphi}
\partial_m\pi) \nn \\
& & - \sqrt{2} {m\over \kappa} \epsilon^{i2}e^{\kappa(\tilde\varphi-{1\over\sqrt{2}}\varphi)}\xi^2 + (\hbox{3-fermion terms}) \nn\\
\delta_\xi \chi^i  &=&  -\I\epsilon^{i j}\sigma^m\bar\eta_j\partial_m\tilde\varphi 
- e^{\kappa\tilde\varphi} \epsilon^{i j}{{\sigma^a}_j}^k   \cD_m\pi^a \sigma^m\bar\eta_k  \nn \\
                 & & +2 {m\over \kappa} \epsilon^{i2}e^{\kappa(\tilde\varphi-{1\over\sqrt{2}}\varphi)}\xi^2+ (\hbox{3-fermion terms})\nn \\
\delta_\xi \pi &=& e^{\sqrt{2}\kappa\varphi}(\I\epsilon_{i j}\Omega^i\xi^j \plushc) \nn \\
\delta_\xi \varphi &=& \epsilon_{i j}\Omega^i\xi^j \plushc \nn \\
\delta_\xi \tilde\varphi &=& \epsilon_{i j}\chi^i\xi^j \plushc \nn \\
\delta_\xi \pi^a &=& e^{-\kappa\tilde\varphi}(-\I\chi^i{{\sigma^a}_i}^j\epsilon_{jk} \xi^k \plushc)
\label{broketrafo} \ ,
}
where 
$\cD_m \pi^a{{\sigma^a}_i}^j = \partial_m\pi^a{{\sigma^a}_i}^j - m({{\sigma^1}_i}^j A_m+{{\sigma^2}_i}^j B_m)$,  ${{\sigma^a}^i}_j = \epsilon^{il}\epsilon_{jk}{{\sigma^a}_l}^k$, 
$\cA_{m}=A_m +\I B_m$, 
and $\tilde\cA_{mn}=\epsilon_{mnop}\cA^{op}$. Here, several field redefinitions have
been performed to facilitate the identification of 4-dimensional $N=2$ multiplets.
With the rescaling $P\rightarrow \sqrt{2/3}P$
the chart of fields in Eqs. (\ref{broke}, \ref{broketrafo}) corresponding to the fields in Eqs. (\ref{five}, \ref{fivetrafo}) is given in Table 6.
\begin{table}[t]
\centerline{
\begin{tabular}{|r|l|} \hline
part. broken $N=2$ $D=4$ SUGRA & $N=4$ $D=5$ SUGRA \\ \hline
$\varphi$  & $(\sqrt{2}\sigma + P)/\sqrt{3}$ \\ 
$\tilde\varphi$   & $(\sigma - \sqrt{2} P)/\sqrt{3}$ \\ 
$\pi$  & $G_{\dot 4}$ \\ 
$\pi^1+\I\pi^2$ & $B_{\dot 4}$  \\
$-\pi^3$  & $ A_{\dot 4}$ \\ 
$\cA_m$  & $C_m$ \\
${e^a}_m$ & ${e^a}_m$ \\
$\psi^1_m$ & $\psi^1_m$ \\
$\psi^2_m$ & $\psi^4_m$ \\
for $i\in\{2,3\} \qquad\matrix{\chi^{4-i} \cr \Omega^{4-i} \cr}$ & $\matrix{\!\!(\lambda^i + \sqrt{2}\nu^i)/\sqrt{3} \cr
\!(\sqrt{2}\lambda^i - \nu^i)/\sqrt{3} \cr}$ \\ \hline
\end{tabular}
}
\caption{Corresponding fields of partially broken $N=2$ $D=4$ supergravity and pure $N=4$ $D=5$ supergravity.}
\end{table}
In terms of $N=2$ multiplets, the fields 
(${e^a}_m$, $\psi^1_m$, $\psi^2_m$, $A_m$) constitute the $N=2$ supergravity multiplet,
($\varphi$, $\pi$, $\Omega^1$, $\Omega^2$, $B_m$) constitute an $N=2$ vector multiplet, and
($\tilde\varphi$, $\pi^1$, $\pi^2$, $\pi^3$, $\chi^1$, $\chi^2$) form a hyper-multiplet.

From the Lagrangian (\ref{broke}) and the supersymmetry transformations (\ref{broketrafo})
it is obvious that $\psi^2_m$ is the gauge field for the spontaneously broken supersymmetry,
whereas a linear combination of $\Omega^1$ and $\chi^1$ is the associated Goldstino. 
Its supersymmetric partner, the scalar $\pi^1+\I\pi^2$ is the Goldstone boson of a spontaneously 
broken central charge gauged by the vector $\cA_m$. That can be seen from
the closure of the first and second supersymmetry algebra on $\pi^1+\I\pi^2$ and $\cA_m$
\cite{np}.

This result should be compared to partially broken 4-dimensional $N=2$ supergravities with complete
$N=2$ multiplets constructed in Refs. \cite{zin, zinovev, fgp-local}. 
Albeit equivalent up to field redefinitions, the non-singular parametrization of the scalar 
fields in Ref. \cite{zinovev} is closest to the parametrization of the scalars as obtained 
by the above-described dimensional reduction, so the field labels from \cite{zinovev} are used
in Table 6. In Ref. \cite{zinovev} the breaking of the $i$th supersymmetry is parametrized 
by the quantities $\mu_i$, $i\in\{1,2\}$.
In this context, the second supersymmetry is chosen to be broken, so $\mu_1=0$ and $\mu_2=m$. The
more general supersymmetry breaking scenario in Ref. \cite{zinovev} can be obtained from five
dimensions by the choice $\mu_1T_2 + \mu_2 T_5$ for the Scherk-Schwarz generator.
One finds complete agreement of (\ref{broke}) and (\ref{broketrafo}) up
to trivial phase factors with the corresponding expressions in Ref. \cite{zinovev}.

\section{Conclusions}
In this paper it was shown that 
partial supersymmetry breaking $N=2 \rightarrow N=1$
in four dimensions can be easily reproduced by compactifying $N=4$ $D=5$ supergravity
on the orbifold $S^1/Z_2$ and using the Scherk-Schwarz mechanism. 
This means that compactification of $N=4$ $D=5$ supergravity on $S^1/Z_2$ automatically
leads to an $N=2$ supersymmetric theory in four dimensions in which no 
prepotential exists for the vector multiplet --- thus allowing for partial supersymmetry 
breaking \cite{fgp-local}. Although the derivation of partially broken theories
in four dimensions from five dimensions is considered here only as a convenient 
tool, it allows for straightforward extensions like matter couplings or embeddings
in higher-$N$ theories in five dimensions. This is possible because the starting point ---
$N=4$ $D=5$ supergravity --- is a linearly realized massless theory with complete multiplets.
Therefore it is
not surprising that complete $N=2$ supermultiplets as in Refs. \cite{zin,zinovev,fgp-local} 
persist in four dimensions in the nonlinearly realized broken phase. 

The Poincar\'e
dualities exploited in Ref. \cite{np} for massless vectors and scalars in four dimensions
can now be seen to be consequences of a duality in five dimension relating a massless
vector to a massless antisymmetric tensor. 
Dual formulations of the theories in Refs. \cite{zin, zinovev, fgp-local} can be found by
dualizing first the Goldstone scalars to antisymmetric tensors and then dualizing
the vectors according to the method described in Ref. \cite{cremdual}. These theories 
correspond to lowest order in the fields to the effective theories derived in Ref. \cite{np}.

In Ref. \cite{ab} the partial breaking of extended supersymmetry in four dimensions with a minimal field
content as dictated by the superHiggs effect was extended to an anti-de Sitter background. There it
was found that only one of the two Goldstone scalars in the theory out of unitary gauge 
could be dualized to an antisymmetric
tensor, the other one could not be dualized because it occurred in the theory without derivatives acting
upon it. If one were to derive this theory by a compactification $AdS_5\rightarrow AdS_4\times S^1/Z_2$,
the resulting 4-dimensional theory would as well contain one vector and one antisymmetric tensor, since
the bosonic group structure of the $N=4$ $AdS_5$ automorphism group requires two of the five vectors,
which reside in the 5-dimensional representation of $USp(4)$ in a Minkowski background, to be replaced
by antisymmetric tensors \cite{romans}. The gauge group then becomes $SU(2)\otimes U(1)$ and the $Z_2$
reflection would project out two of the vectors in $SU(2)$ and one antisymmetric tensor in $U(1)$. 
Work along these lines is in progress. 

\section*{Acknowledgments}
I would like to thank the FNAL theory group for their hospitality during my stay as
a summer visitor, when this work was initiated. I am also grateful to
J.~Bagger and Yi-Y.~Wu for helpful discussions.

\section*{Appendix}
\begin{appendix}

The  5-dimensional Dirac algebra with the space-time metric $\eta_{AB} = \diag(-1, 1, 1, 1, 1)$ reads
$$
\{ \Gamma^A, \Gamma^B \} = -2\eta^{AB} \ ,
$$
where $\Gamma^A \in \{ \gamma^0, \gamma^1, \gamma^2, \gamma^3, \gamma^5  \}$.
The matrices $\gamma^a$ and $\gamma^5$ are defined as in Ref. \cite{wb} and the antisymmetric combinations
$\Gamma^{A_1\dots A_n} = {1\over n!}\Gamma^{[A_1}\cdots\Gamma^{A_n]}$ are defined with strength one. 
The 5-dimensional epsilon-tensor is defined by $\epsilon^{01234} = 1$.

The real symplectic metric to lower $USp(2N)$-indices is chosen to be $\Omega_{ij} = 1_N \otimes \I\sigma^2$.
It is used to raise and lower symplectic indices of the vectors $V^i$ and $V_i$ according to 
$V^i =\Omega^{ij}V_j$ and
$V_i =\Omega_{ij}V^j$.

A symplectic Majorana spinor in five dimensions (denoted by upper case Greek letters) 
is written in terms of Weyl spinors in four dimensions (denoted by lower case Greek letters) as
$$
\Psi^i = \pmatrix{ \psi^i_\alpha \cr -\Omega^{ij}\bar\psi_j^{\dot\alpha} \cr}\ ,
\ \ \bar\Psi_i = 
(\Omega_{ij}\psi^{j\alpha}\ \ \bar\psi_{i\dot\alpha})\ ,
$$
where $\bar\psi_i^{\dot\alpha} = \epsilon^{\dot\alpha\dot\beta}(\psi^i_\beta)^*$. ``Symplectic'' indices
on Weyl spinors in four dimensions are merely labels --- they are not covariant indices.

The massless superalgebra in five dimensions has a $USp(N)$ automorphism group. 
$USp(N)$ with $N$ even is the compact Lie group defined by the set of complex $N\times N$ matrices
that are both unitary and symplectic. The generators of the corresponding Lie algebra $usp(N)$ form a set of $N(N+1)/2$
hermitian matrices $T_r$ that satisfy the symplectic condition $T_r \Omega_{..} + \Omega_{..} T_r^{\top} = 0$
$\quad\forall r \in \{1,...,N(N+1)/2\}$.
The representation of the basis elements of $usp(4)$ used in this paper is 
$$
T_r \in \left\{ \pmatrix{\sigma^j & 0 \cr 0 & 0 \cr}, \pmatrix{0 & 0 \cr 0 & \sigma^j \cr},
\pmatrix{0 & \sigma^1 \cr \sigma^1 & 0 \cr},
\pmatrix{0 & -\sigma^2 \cr -\sigma^2 & 0 \cr},\pmatrix{0 & \sigma^3 \cr \sigma^3 & 0 \cr},
\pmatrix{0 & \I \cr -\I & 0 \cr}
\right\}
$$
with $r \in \{1,...,10\}$.
\end{appendix}





\end{document}